\documentclass[a4paper,english,10pt,twoside]{article}
\usepackage[english]{babel}
\usepackage[latin1]{inputenc}
\usepackage{amsmath}
\usepackage{graphicx}

\usepackage{fancyhdr}
\fancypagestyle{plain}{%
  \fancyhf{}%
  \fancyhead[C]{\small\it{Twenty years of giant exoplanets - Proceedings of the Haute Provence Observatory Colloquium, 5-9 October 2015\\ Edited by I. Boisse, O. Demangeon, F. Bouchy \& L. Arnold} }
  \fancyfoot[C]{\normalsize \thepage}%
}

\newcommand{\nc}   {\newcommand}
\nc{\dsfrac}[2]{{\displaystyle\frac{#1}{#2}}}

\newcommand{\Jcal}  {{\EuScript J}}

\nc{\ergs}  {\mathrm{erg}/\mathrm{gm\;s}}
\nc{\ergc}  {{\rm erg}~{\rm cm}^{-3}}
\nc{\ergcs} {{\rm erg}~{\rm cm}^{-2}~{\rm s}^{-1}}
\nc{\ms}   {{\rm m}~{\rm s}^{-1}}

\nc{\Eem}   {\widetilde{E}_\mathrm{e}}
\nc{\Eexm}  {\widetilde{E}_\mathrm{ex}}
\nc{\Eim}   {\widetilde{E}_\mathrm{I}}
\nc{\ER}    {E_\mathrm{R}}
\nc{\FL}    {F_\mathrm{Ly}}
\nc{\FR}    {F_\mathrm{R}}
\nc{\FRa}   {F_\mathrm{R,1}}
\nc{\FFR}   {\mathbf{F}_\mathrm{R}}
\nc{\Fe}    {F_\mathrm{e}}
\nc{\FFe}   {\mathbf{F}_\mathrm{e}}
\nc{\Fsat}  {q_\mathrm{sat}}
\nc{\Ke}    {\EuScript{K}_\mathrm{e}}
\nc{\mfp}   {l_\mathrm{e}}
\nc{\mrc}   {\mathrm{c}}
\nc{\mrr}   {\mathrm{r}}
\nc{\nH}    {n_\mathrm{H}}
\nc{\nel}   {n_\mathrm{e}}
\nc{\Pe}    {P_\mathrm{e}}
\nc{\Pg}    {P_\mathrm{g}}
\nc{\PR}    {P_\mathrm{R}}
\nc{\Qelc}  {Q_\mathrm{elc}}
\nc{\Qinc}  {Q_\mathrm{inc}}
\nc{\gyr}   {r_\mathrm{B}}
\nc{\Ta}    {T_a}

\nc{\divFe} {\nabla\cdot\FFe}
\nc{\divFR} {\nabla\cdot\FFR}
\nc{\FLj}   {F_\mathrm{Ly,\Jcal}}
\nc{\FRj}   {F_\mathrm{R,\Jcal}}

\nc{\aap}   {A\&A}
\nc{\aaps}   {A\&AS}
\nc{\apj}   {ApJ}
\nc{\mnras}   {MNRAS}
\nc{\aapr} {A\&ARv}
\nc{\pasp} {PASP}
\usepackage{txfonts}
\usepackage[squaren,cdot]{SIunits}% SI units of measurement (\centi\meter)
\usepackage{tabularx}
\usepackage[breaklinks=true]{hyperref} %% to avoid \citeads line fills
\usepackage{natbib,twoopt}
\bibpunct{(}{)}{;}{a}{}{,} %% natbib format for A&A and ApJ
\makeatletter
\newcommandtwoopt{\citeads}[3][][]{\href{http://adsabs.harvard.edu/abs/#3}%
{\def\hyper@linkstart##1##2{}%
\let\hyper@linkend\@empty\citealp[#1][#2]{#3}}}
\newcommandtwoopt{\citepads}[3][][]{\href{http://adsabs.harvard.edu/abs/#3}%
{\def\hyper@linkstart##1##2{}%
\let\hyper@linkend\@empty\citep[#1][#2]{#3}}}
\newcommandtwoopt{\citetads}[3][][]{\href{http://adsabs.harvard.edu/abs/#3}%
{\def\hyper@linkstart##1##2{}%
\let\hyper@linkend\@empty\citet[#1][#2]{#3}}}
\newcommandtwoopt{\citeyearads}[3][][]%
{\href{http://adsabs.harvard.edu/abs/#3}
{\def\hyper@linkstart##1##2{}%
\let\hyper@linkend\@empty\citeyear[#1][#2]{#3}}}
\makeatother

\makeatletter 

\def\@maketitle{%
  \vskip 2em%
  \begin{center}%
  \let \footnote \thanks
    {\LARGE\textbf \@title \par}%
    \vskip 1.5em%
    {\normalsize
      \lineskip .5em%
      \begin{tabular}[t]{c}%
        \@author
      \end{tabular}\par}%
    \vskip 1em%
    {\normalsize \@date}%
  \end{center}%
  \par
  \vskip 1.5em}
\makeatother

\newcommand{\affil}[1]{\small{\hskip-0.55cm #1}}
\newcommand{\acknowledgments}[1]{\small{ \vskip3mm \hskip-0.55cm {Acknowledgments: #1}}}

\begin{document}

\setcounter{page}{1}  % The final page number will be set by the editors. First page number of a paper is always odd.

\title{K-Stacker, a new way of detecting and characterizing exoplanets with high contrast imaging instruments} 
\author{Le Coroller $^1$ $^2$, H., Nowak$^2$, M.,  Arnold$^{2}$, L., Dohlen$^1$, K., Fusco$^{1,3}$, T.,  Sauvage$^{1,3}$,  J.F., Vigan$^1$, A.}
\date{} %none required, but you can add the date of your talk or the date you finished the writing of your paper.
\maketitle
\affil{  $^1$Laboratoire d'Astrophysique de Marseille, 38 rue Fr\'ed\'eric Joliot-Curie, \\13388 Marseille Cedex 13, France (\texttt{herve.lecoroller@lam.fr})\\
          $^2$Aix Marseille Université, CNRS, OHP (Observatoire de Haute Provence), Institut Pythéas UMS 3470, 04870 Saint-Michel-l'Observatoire, France \\
          $^3$ Onera, The French Aerospace Lab, 92322 Châtillon, France
}

\vskip1cm

\begin{abstract}
This year, a second generation of coronagraphs dedicated to high-contrast direct imaging of exoplanets is starting operations.
Among them, SPHERE, installed at the focus of the UT3 Very Large Telescope, reaches unprecedented contrast ratios up to $10^{-6}$ -$ 10^{-7}$,
using eXtreme Adaptive Optics and the Angular Differential Imaging (ADI) techniques.
 In this paper, we present a new method called Keplerian-Stacker that improves the detection limit of high contrast  instruments like SPHERE, by up to a factor of 10.
It consists of observing a star on a long enough period to let a hypothetical planet around that star  move along its orbit.
Even if in each individual observation taken during one night, we do not detect anything, we show that it is possible, using an optimization algorithm, to re-center the images according to keplerian motions (ex: 10-100 images taken over a long period of typically 1-10 years)
and detect planets otherwise unreachable. This method can be used in combination with the ADI technics (or possibly any other high contrast data reduction method) to improve the Signal to Noise Ratio in each individual image, 
and to further improve the global detection limit. It also directly provides orbital parameters of the detected planets, as a by-product of the optimization algorithm.

\end{abstract}

\section{Introduction}

\begin{figure}[h]  % figure here
  \centering
  \includegraphics[height=6cm,width=12cm]{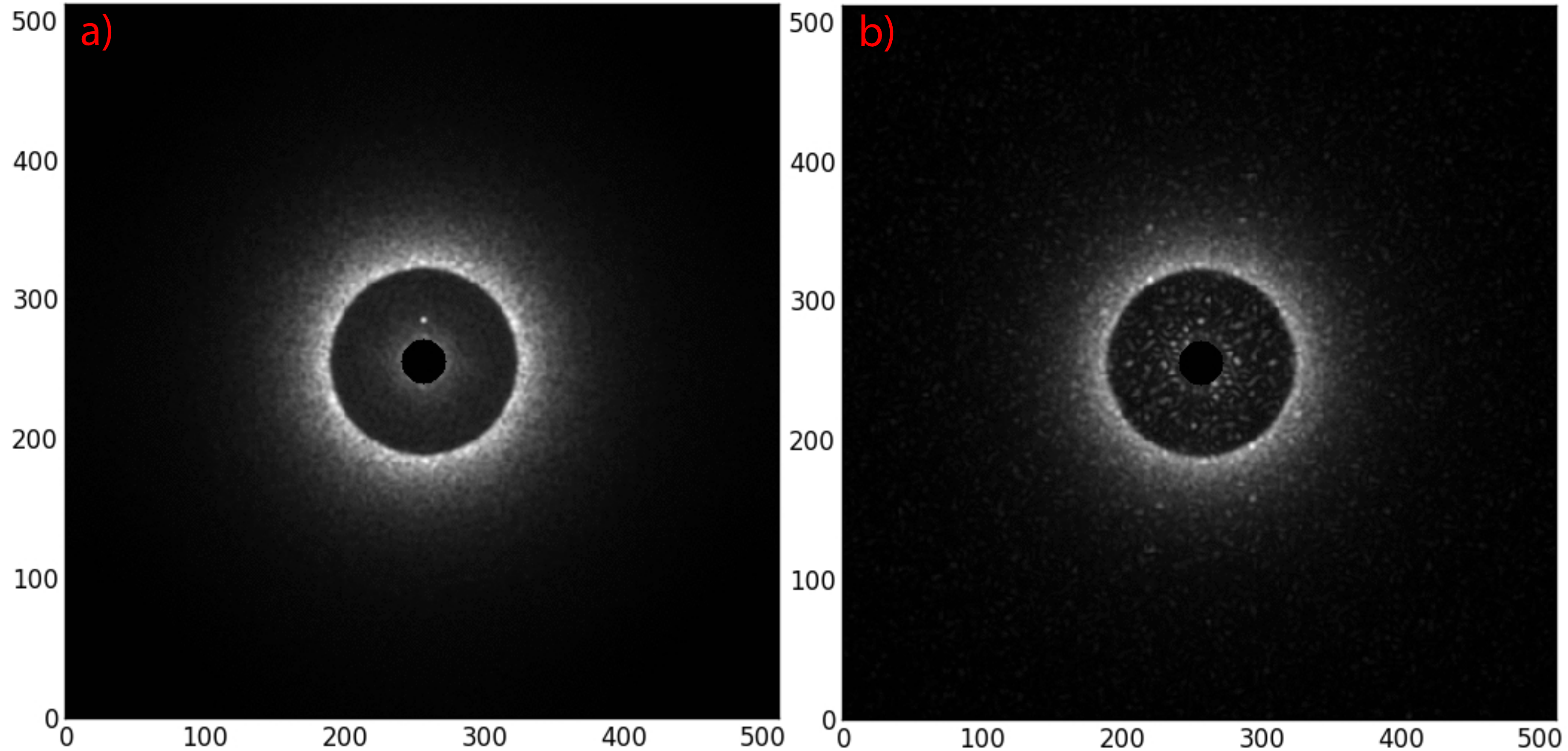}
  \caption{Simulation of images obtained with the Sphere/IRDIS instrument (ALC2 configuration) at 1.6 micron. 
  The phase masks at the output of the XAO have been simulated using \citetads{2006OExpr..14.7515F} code. 
  A planet is introduced in these images just above the coronagraphic mask.
  a) one second of exposure time without static speckles. b) Same simulation than in a) but we have added fixed speckles. 
  The planet is un-detectable with the static speckles. The ADI technique consists by removing these static speckles (no ADI simulations shown here).
  The central part has been masked out to increase visibility.}
  \label{speckles}
\end{figure}

\begin{figure}[h]  % figure here
  \centering
  \includegraphics[width=12cm]{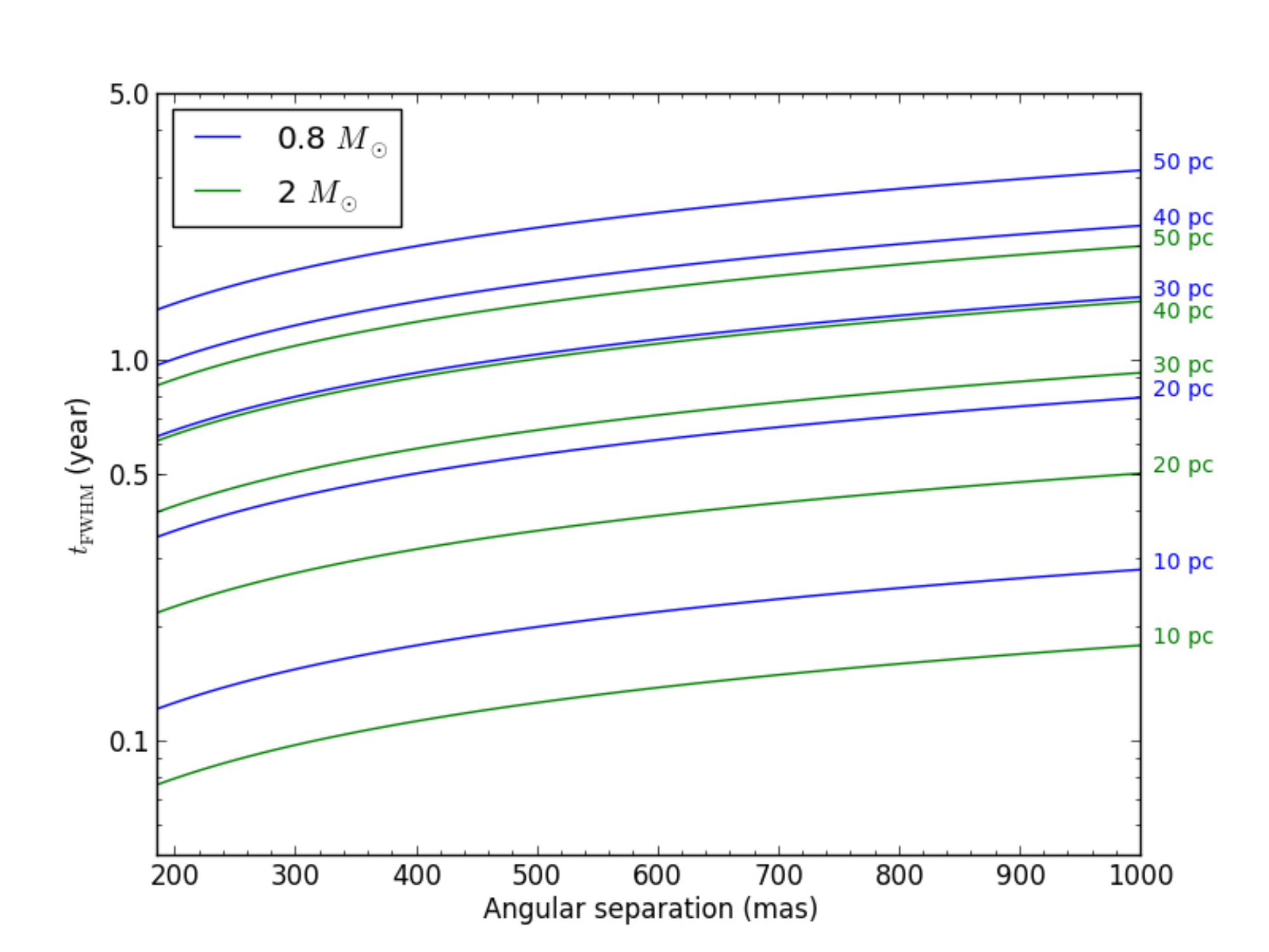}
  \caption{This plot shows the typical time (in year) it takes for a planet orbiting a star between 10 and 50 PC away to move of one FWHM PSF. 
  The size of the PSF is defined for a telescope of $8\, \meter$ observing at $1.6\, \micro \meter$.
  The horizontal axis gives the separation of the planet from the star, view from the earth in milli-arcseconds.
  The curves have been plotted for a star of 0.8 (blue) and 2 (green)  solar masses. 
  To simplify and simply to get an order of magnitude of the expected motions, we have supposed only circular orbits perpendicular to the line of sight.}
  \label{PSF}
\end{figure}

Most of the 1960 exo-planets detected until now have been found using indirect methods; in particular radial velocity technique and  photometric transits.
Indeed, it is extremely difficult to see the planet light that is drowned in the diffracted light of the star. 
A Jupiter and an Earth like planets are about $10^{-8} - 10^{-10}$ fainter than their parent star!
Nevertheless, huge progress have been done these last twenty years with Adaptive Optics and Coronagraphic systems in order to remove the light of the star and to be able to detect directly the  light of the planets. This last two years, two new instruments SPHERE \citepads{2015Msngr.159....2L,2008SPIE.7014E..18B} and GPI \citepads{2014PNAS..11112661M}, equipped with last technologies such as an eXtreme Adaptive Optics system and an apodized coronagraph have started their scientific observations. For the first time these instruments are able to reach a contrast of $10^{-6}$ and can detect young Jupiter-like planets,  equivalent to the giant planets in our own solar system but formed recently. SPHERE and GPI are also equipped with a low resolution spectrograph that allows to get spectrum of the atmosphere of these planets. In order to reach very high contrast  ($10^{-6}$), the stars are observed close to the transit to take advantage of the maximum field rotation in the Angular Differential Imaging technique \citepads{2006ApJ...641..556M}. The ADI method consists in letting the Field of View of an altitude/azimuth telescope rotates while keeping the instrumental optics fixed. During such an observation, the planets move with the FoV while the quasi-static speckles stay fixed in the focal plan of the instrument. It is then possible to create a reference PSF (a "map" of the fixed speckles without the planet), for example by taking the median of the images. Subtracting this reference PSF from each image then increases the contrast (see Fig. \ref{speckles}).
Several mathematical techniques have been proposed to create this reference PSF \citepads{2006ApJ...641..556M,2007ApJ...660..770L,2012MNRAS.427..948A}, which allow a gain in contrast of 10-100 for ADI observations of about 1 hour of total exposure time.
But, no matter which mathematical method is employed, the ADI technique cannot be used with exposure times longer than 1-2 hours because far away from the transit, the field rotation is too slow (to create a reference PSF without the planet, the planet has to move of more than one PSF Full Width at Half Maximum during the exposure time). It is also not possible to add several ADI images taken over several months or years because the planet moves on its Keplerian orbit (Fig. \ref{PSF}). Note that we can observe the same star one hour each night (at the transit, when ADI technique works) during one run of 7-10 days to reach 7-10 hours of total exposure time. But, such an observation strategy would consume a lot of time to increase just a little (maximum by a factor of three in this example) the accessible contrast limit.   Indeed, even if we detect new objects, we will have to re-observe at least 2-3 more times (14-30 h minimum!) to know if these detections are background stars or linked objects. Moreover, we cannot be sure that the speckles remaining in ADI images taken over only a few days are well decorrelated... Thus, nothing tells that the SNR will increase as the square root of the total exposure time. \\

\noindent  In this paper, we present a new method, named Keplerian-Stacker, to detect exoplanets. This technique allows to find the orbital parameters of planets that we do not see in a serie of individual images taken over several months! It is then possible to recentre the images and co-add them to increase the contrast limit accessible with an instrument like SPHERE. K-Stacker could allow to detect planets impossible to see in only 1-2 hours of ADI exposure time. \\
\noindent In Sect. \ref{principal}, we describe the principle of K-Stacker. In Sect. \ref{simulations}, we show an encouraging simulation demonstrating that it is possible to find the orbital parameters in about 10 hours of computation using a modest cluster of 190 processors. A discussion and conclusions are given at Sect. \ref{conclusion}.

%----------------------------------------------------------------
%----------------------------------------------------------------

%----------------------------------------------------------------
%----------------------------------------------------------------

\section{K-Stacker principle}
\label{principal}

The idea of K-Stacker is to determine the correct orbit of a planet hidden in a set of individual images taken over several months. To do so, an algorithm tries many different orbits evaluates the following SNR function (Eq. \ref{SNR}) for each of the Keplerian motion test, and optimizes it.

\begin{equation}
SNR(t_0,e,a,\theta_0,\Omega,i,M_*,d)=\frac{\sum_{i=1}^N {F_i}}{\sigma}
\label{SNR}
\end{equation} 

with $\sigma=\sqrt{\sigma_1^2+\sigma_2^2+...+\sigma_N^2}$\\

SNR is a function of 8 parameters. The classical 7 parameters that are required to define a Keplerian motion, plus the distance of the star used to know the size of the projected ellipse on the final CCD:

\begin{itemize}
\item $a$ : semi-major axis
\item $e$ : eccentricity
\item $t_0$ : time at the perihelion passage (in our simulations $t=0$ at the first observation)
\item $M_*$ : star mass
\item $\Omega$ : longitude of the ascending node
\item $i$ : inclination
\item $\theta_0$ : argument of periapsis (orientation of the ellipse in the orbital plane)
\item $d$ : distance of the star\\
\end{itemize}

\begin{figure}[h]  % figure here
  \centering
  \includegraphics[width=12cm]{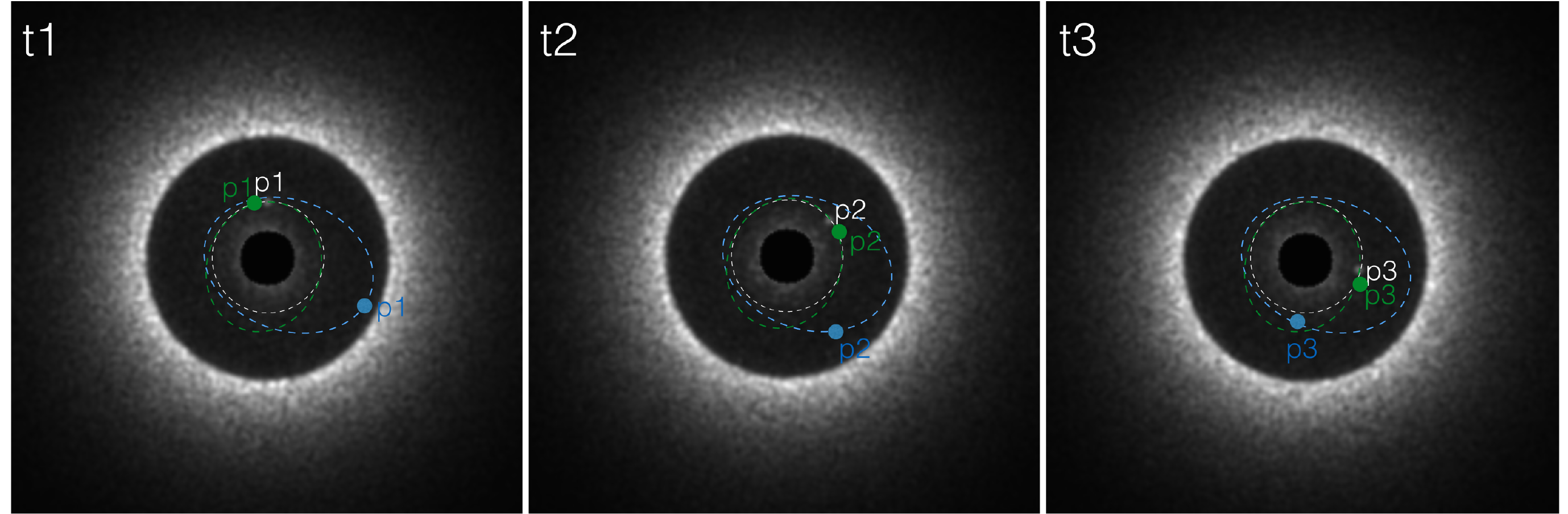}
  \caption{Principle of Keplerian-Stacker: A planet is moving in these three images taken over several months at three different dates t1, t2, t3. 
 The planet is at the position P1, P2, P3 in white. To find this solution, K-Stacker algorithm tests a lot of orbits. For each orbit, it computes Eq. \ref{SNR}. In this example, the blue orbit is totally wrong while the green orbit is closer to the correct solution. With the green orbit, the SNR value (Eq. \ref{SNR}) is slightly better because a part of the planet flux has been integrated (the green circles slightly overlap the planet in images at times t1 and t2). K-Stacker has found the planet when the SNR reaches its maximum value i.e. the circles are superimposed to the planet position in each image. To give a simple example, we have shown only three orbits (white, blue and green), and only three images, in which we can actually see the planet (in white at the position p1, p2, p3). But the idea of K-Stacker is to use 10-100 images taken over several months or years and to search for planets that we do not see in any individual image. We have found that  $>10^8$ orbits have to be tested to find the planets in the field of view of the corrected area by the XAO of the SPHERE/IRDIS instrument.}
  \label{principe}
\end{figure}

For a set of  orbital parameters, the position of the planet in each image i, taken at a specific date is perfectly known.
$F_i$ is the flux integrated at the position of the planet in the image i, in a circle of the size of the instrumental PSF, and corrected for the sky background.
$\sigma$ is the root square of the quadratic sum of the standard deviations computed around each circle (theoretical position of the planet) in each image i (see Fig. \ref{principe}). 
For a set of tested orbital parameters ($a$, $e$, $t_0$, $M_*$, $\Omega$, $i$, $\theta_0$), SNR simply gives the signal to noise that we should obtain at the final position of the planet if we add the N images (observations) recentered with those parameters.

%----------------------------------------------------------------
%----------------------------------------------------------------

In our simulations, we assume that the mass and the distance of the star are known by others observations (HIPPARCOS, GAIA, spectroscopic observations, etc.).
We have then to find six parameters. We have shown that about $10^8$ orbits have to be tested to find the planets in the field of view of the corrected area by the XAO of the SPHERE/IRDIS instrument \citep{mathias_preparartion}.

\section{Simulations and first results}
\label{simulations}

\subsection{Coronagraphic simulations}
\label{corono}

We have simulated images of the SPHERE/IRDIS instrument (configuration ALC2). These simulations have been done at $\lambda=1.6\, \micro \meter$, with a fried parameter of $r_0=0.8$, for a star of magnitude $R=8$ (used for the performance of the AO only), and using a code developed by \citetads{2006OExpr..14.7515F} to create the XAO residual phase error masks. Each long exposure (see fig. \ref{figsimulation}) is the sum of one hundred images of about 10 ms (frozen atmosphere). We have not calibrated our images in flux by taking into account the star magnitude.

\subsection{K-Stacker simulations}

Following the method described in Sect. \ref{corono}, we have simulated 100 SPHERE/IRDIS coronagraphic images of 1 second of exposure time. Assuming that these images have been taken regularly during 3 years (i.e. one observation every 11 days), we then introduced a planet below the detection limit in each individual image, at the position defined by a set of randomly chosen orbital parameters. We then used K-Stacker to retrieve these parameters.

\subsection{The optimisation algorithms and the computation time}

To be of any practical interest, an algorithm such as K-stacker has to be able to find the orbital parameters in a reasonable amount of time. To this aim, we have developed an algorithm based on a combination of a brute-force search, and a local gradient optimization algorithm. The grid used for the brute-force search is the result of a difficult trade-off between reliability and time consumption, and its construction will be more extensively discussed in a forthcoming paper. We found that about $10^8$ points have to be used in order for the algorithm to be reliable enough. With such a grid, several SNR maximums can be mistaken for the correct solution, and we cannot find the orbital parameters with a good accuracy. To converge more accurately toward the correct orbital parameters, we added a second stage to the algorithm, in which some of the best solutions found over the grid are re-optimized using a "local" gradient algorithm that searches the closest maximum. Again, the total number of solution to be re-optimized has to be carrefully chosen, and will be discussed in \cite{mathias_preparartion}.

\begin{figure}[h]  % figure here
  \centering
  \includegraphics[width=12cm]{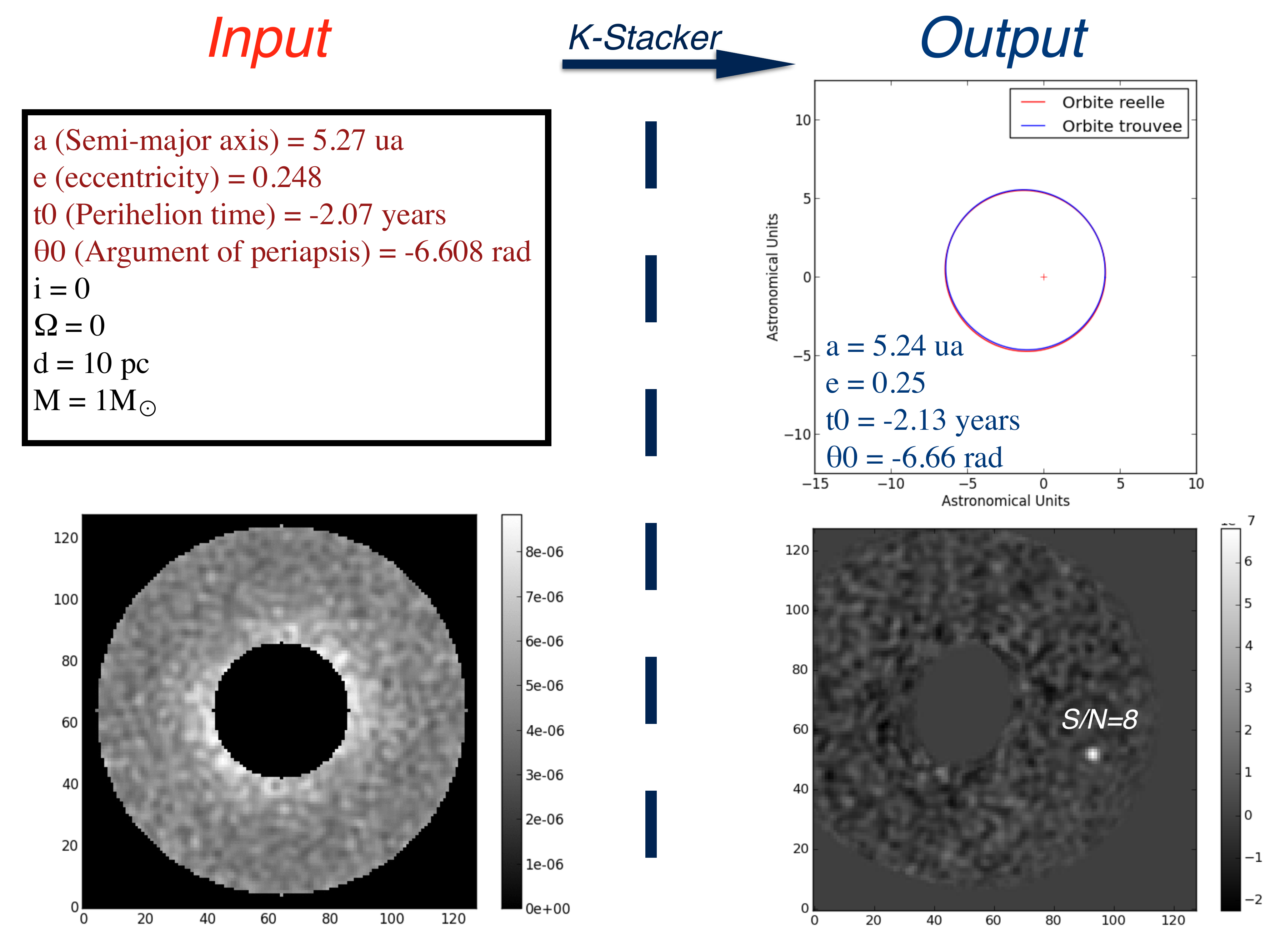}
  \caption{Top-Left: Orbital parameters of the planet. Bottom-Left: one of the one hundred "long exposure" ($1s$) coronagraphic images taken regularly over 3 years. A planet is moving in these images with the orbital parameters given above (Top-Left). In each image, the signal to noise of the planet is of about 0.8 and it is absolutely undetectable! Top-right: orbital parameters found by K-Stacker. To limit the total computation time (on one processor), in this simulation we have searched for only four parameters: $a$, $e$, $t_0$ and $\theta_0$. $i$ and $\Omega$ are fixed at zero i.e. the orbit is perpendicular to the line of sight. The star's mass and distance are supposed to be known by other means. Bottom-right: the one hundred images have been re-centred using the orbital parameters found; each image is rotated in order to put the planet on the perihelion axis, and then it is translated to superimpose the flux of the planet of all the observations. We perfectly detect the planet and we show that the signal to noise has increased like the square root of the total exposure time.}
  \label{figsimulation}
\end{figure}

In order to limit the computation time, in the first simulation presented in Fig. \ref{figsimulation}, we have searched for only four parameters ($a$, $e$, $t_0$, $\theta_0$). This limits the number of orbits in the 4-dimension grid search to $\approx 10^4$, and makes possible to test the algorithm in various conditions quickly. $i$ and $\Omega$ have been fixed at zero (the orbit is seen face-on), and we have done the assumption that the mass and the distance of the star are known by other observations. We have been able to find the correct solution (right panel of Fig. \ref{figsimulation}) in only a few hours with one processor. By parallelizing the computation on the 16 nodes x 12 processors ($\approx 192$ times faster than on one processor) of the LAM computing cluster, K-Stacker should be able to find the six parameters in less than 10 hours. Note, that clusters at least 10-100 times more powerful than the LAM one exist, and could be used for increased performances. This could allow to let the algorithm also optimize the mass and the distance of the central star, when they are known with poor accuracy.

Given these results, it appears that K-Stacker can find the 7 orbital parameters of a planet moving in 25-100 images of the A.O. corrected area  (1.6 arcsec in diameter) of a SPHERE/IRDIS instrument in a reasonable computation time. It is even probably possible to search for planets beyond this corrected area but more study is required to determine the limit defined by the maximum feasible computation time. Note that a Brute-Force algorithm tests all the possible orbits. Thus, It should not take more time to find one planet than several (around the same star). For example, for three planets, K-Stacker will obtain three SNR higher than 5 among the $10^8$ tested orbits, corresponding to the three sets of orbital parameters. A more detailed study on the required computating resources will be given in a forthcoming paper \citep{mathias_preparartion}.
 
%----------------------------------------------------------------
%----------------------------------------------------------------

\section{Discussion and Conclusion}
\label{conclusion}

We have already done several tests to search for planets with various orbital parameters. Recently, we have even been able to constrained six orbital parameters \citep{mathias_preparartion}. Each time that the total SNR on the recentered images was higher than 5 in the total recombined image,
we have been able to detect the planet and to confirm that we have found the correct orbital parameters, even if the signal to noise ratio was under one in each individual image ! 
In this first work, we do not simulate the instrumental pseudo-static speckles, and how the SNR will increases if we add (after we have re-centered) the images reduced by the ADI technique. 
Nevertheless, the speckles of images taken over several months or years should be very well decorrelated. But, if pseudo-static speckles are still present after ADI reduction, it is probably possible to use a Keplerien-ADI method to remove them. Where ADI uses the parallactic rotation to differentiate speckles from planets, Keplerian-ADI instead uses the orbital motion such as proposed by \citetads{2015arXiv151002478M} with their ODI technique. But, Keplerian-ADI can be used with any set of orbital parameters (long, or short periods, eccentric, etc.) even if we don't see the planet in each individual images. By re-centering the images, the static speckles should also be very well scrambled; more studies will be required on the true data. We are currently testing K-Stacker in various situations
 (various orbital parameters; inhomogeneous data with various seeing; etc.) to check if we can propose a very robust algorithm able to detect planets in any conditions. 
K-Stacker should allow to obtain very long exposure times by summing $10-100$ images taken over several months/years in order to detect fainter planets, down to a contrast of $10^{-8}$ with an instrument such as SPHERE. 
Saturn and Neptune like planets could be detectable with SPHERE using K-Stacker. K-Stacker could also be used to split one observation in several shorter ones. For example, it is more interesting to do six observations of ten minutes 
rather than one observation of one hour, because in both cases we detect the planet (after the K-Stacker recombination in the first case) at the same level of signal to noise but with the first observing strategy (splitting in six observations), we also get the orbital parameters.
Nevertheless, if we want to use K-Stacker with ADI-reduced images, it has to be noted that the minimum exposure time of one observation is defined by the required time for the planet to move of one PSF due to the field rotation. 
Therefore, with SPHERE, this strategy (splitting in several short observations) will be possible only for stars at high altitude (with faster Field rotation at the transit).
 It will be possible to do shorter exposure times with the E-ELT even using the ADI method, because the PSF size of the 40 m will be five times smaller than the PSF of the VLT. 
 It will then be perhaps possible to add several ADI images of short exposure time (ex: $10-15$ min) on a lot of stars (at various altitudes). 
Thus, K-Stacker can be a very efficient method to minimize the total exposure time required to detect planets and find their orbital parameters. 
More studies are required to be sure that we will be able to find the orbital parameters in a reasonable computation time with the  E-ELT Field Of View (FoV corrected by the XAO).
It also depends on how powerful computer clusters will be in about ten years. To use K-stacker, the center of the images must also been known with an accuracy of about 0.1 FWHM-PSF. It is the case on SPHERE using the waffles.
On the E-ELT, K-Stacker could allow to detect Earth-like planets. A more detailed description of the K-Stacker algorithm, and a discussion of the most recent results obtained will be given in a forthcoming paper \citep{mathias_preparartion}.
  
\acknowledgments{We thank the organizers of the OHP 2015 colloquium for this wonderful conference.}

%****************************************************************************************
%****************************************************************************************
%****************************************************************************************
%****************************************************************************************

%-------------------------------------------------------------------

\bibliography{lecoroller_biblio}   %>>>> your bibliography data/list in this authorname_biblio.bib file
\bibliographystyle{aa} % style aa.bst

\end{document}